# Contextualization of Big Data Quality: A framework for comparison

MOSTAFA MIRZAIE, Ferdowsi University of Mashhad (FUM), Mashhad, Iran
BEHSHID BEHKAMAL, Ferdowsi University of Mashhad (FUM), Mashhad, Iran
SAMAD PAYDAR, Ferdowsi University of Mashhad (FUM), Mashhad, Iran

With the advent of big data applications and the increasing amount of data being produced in these applications, the importance of efficient methods for big data analysis has become highly evident. However, the success of any such method will be hindered should the data lacks the required quality. Big data quality assessment is therefore a major requirement for any organization or business that use big data analytics for its decision making. On the other hand, using contextual information is advantageous in many analysis tasks in various domains, e.g. user behavior analysis in the social networks. However, the big data quality assessment has benefited less from this potential. There is a vast variety of data sources in the big data domain that can be utilized to improve the quality evaluation of big data. Including contextual information provided by these sources into the big data quality assessment process is an emerging trend towards more advanced techniques aimed at enhancing the performance and accuracy of quality assessment. This paper presents a context classification framework for big data quality, categorizing the context features into four primary dimensions: 1) context category, 2) data source type that contextual features come from, 3) discovery and extraction method of context, and 4) the quality factors affected by the contextual data. The proposed model introduces new context features and dimensions that need to be taken into consideration in quality assessment of big data. The initial evaluation demonstrates that the model is more understandable, more comprehensive, richer, and more useful compared to existing models.

## 1   INTRODUCTION

The increasing developments in the information and communication technology (ICT) domain during the last decade has given rise to the big data applications, mainly manifested in the enormous volume of data being produced or consumed in business applications [1]. The data volume generated by big data applications is projected to grow from 2.7 Zettabytes in 2012 to 35 Zettabytes by the year 2020 [2], [3]. The large volume of data being quickly generated in these applications can set the stage for new issues, one of which is the lack of efficient and effective quality control techniques, leading to poor data quality [4]. This can affect the results obtained from analysis of the big data, leading organizations to make the wrong business decisions [5].

In the absence of effective data quality management techniques, organizations need to spend more time budget to reconcile data, causing delay in the development of new applications, loss of credibility, and compliance problems, all resulting in excessive cost [6]. IBM has stated that 33% of corporate executives do not trust their processing results [7]. Also, Redman [8] also mentions that an annual cost of three trillion dollars is imposed on USA due to poor data quality. Consequently, big data quality management, which entails quality assessment, needs to be considered to reduce organizational costs and improve the quality the decisions made based on big data.

Recently, numerous research publications have focused on the big data quality. Based on the results of our recently conducted systematic review on big data quality [9], we have found that most of the studies aim at monitoring and detection of poor data quality, while fewer investigate quality assessment or improvement. On the other hand, only few of these studies consider contextual information in the process of data quality assessment. In the context-based techniques, quality of data is determined not only through analysis of local application-specific information, but also using information from a global context, which in turn enhances the performance of big data quality management, and solves issues like concept drift [10].

Another study emphasizes the use of ontologies as rich sources of information for data cleaning and uses contextual information to improve the accuracy of data quality management [11]. Considering the existing works in this area, the important question is whether any other context features can improve the big data quality management process. Given the importance of managing big data quality with the help of contextual information, and the fact that no context model has been presented in this area, we are going to identify and classify contextual data that can be used for quality management of big data. To highlight the need for context awareness in data quality assessment, here, we will present a scenario in healthcare domain in which poor data quality can introduce strong threats resulting in catastrophic consequences, e.g. patient's death.

In a health monitoring system, various vital signs of a patient might need to be continuously read by the sensors and monitored via a body sensor network. The monitoring results are then sent to a remote system in a hospital or clinic. Data quality problems can arise in several steps during generation, communication, and consumption of this streaming sensory data. Considering the lifetime of the patient's vital signs data, it is possible that the reason for low data quality is the sensor's lack of required precision or its hardware limitations. Any such constraint in the components of the system, e.g. the sensors, can lead to poor quality of the output data. As a result, being context-aware implies that the system should take these constraints into consideration in the data quality assessment process. The required information on this matter may not be provided internally by the system, e.g. through detailed specification of the sensors in the system, making it necessary to obtain this information from specialized and reliable external data sources such as Vernier[1]. This emphasizes the need for considering both internal and external sources for contextual data. Getting further from the source of data generation, it is possible that external or environmental factors can bring about quality issues in the data. For instance, weather conditions may affect the precision of the sensors and result in low quality of data. Therefore, to assess the quality of the data produced by the sensor, due care needs to be given to information about the environmental context. Additionally, it is possible that some of the data quality problems are due to organizational constraints such as insufficient budget for resource allocation or under skilled staff responsible for data processing. Having precise information about the organizational constraints or immaturities can improve the big data quality management process. This information can be obtained from the corresponding organization or from other available reference organizations that publish status and competency reports about the organizations.

Considering the points mentioned, it is evident that a great deal of information needs to be considered under the umbrella of contextual features to provide the required information for a big data quality management process and improve its performance. Our systematic review of the literature [9] demonstrates that different aspects of the contextual information has been overlooked in the field. This has motivated the current study with the purpose of providing a comprehensive context model for big data quality.

The rest of the paper is organized as follows: Section 2 provides different definitions of context mentioned in the literature, and in addition, briefly discusses context-aware big data studies. The proposed context model is introduced in Section 3, and it is used in Section 4 to compare existing big data quality studies. Section 5 is dedicated to the evaluation of the proposed framework, and finally, Section 6 concludes the paper.

---

[1] https://www.vernier.com/products/sensors/bps-bta/

## 2 LIERATURE REVIEW

In this section, we first review different definitions of context that is presented in the literature. Then, we will discuss the context-aware methods proposed in big data research.

### 2.1 Context Definition

Despite its common usage, "context" is used in different disciplines to mean different things. In Wikipedia, it is defined as "*a frame that surrounds the event and provides resources for its appropriate interpretation*". On the other hand, in Merriam-Webster's Collegiate Dictionary[2], it is defined as "*the interrelated conditions in which something exists or occurs*".

Some researchers have provided their own definition of context, the first of which is in 1994 by Schilit [12] who defines context as the location, identities of an object, and its changes. Later in [13] he improves his own definition by stating that "*three important aspects of context are: where you are, who you are with, and what resources are nearby*". Although in this definition the information of the surrounding sources is taken into account, the definition of context is still mainly based on location.

In 1999, Schmidt [14] reported that context is "*knowledge about the user's and IT device's state, including surroundings, situation, and to a less extent, location*". In this study, context is very specific and only includes user and device information, while other important information can also be considered as context.

In the same year, Abowd [15] presents a task relevant definition and holds the view that context is "*any information that can be used to characterize the situation of an entity. An entity is a person, place, or object that is considered relevant to the interaction between a user and an application, including the user and applications themselves*". This definition is also presented for the computing application domain and cannot be considered the best definition for big data quality.

Chen [16] argues that the above definitions are not suitable for mobile computing and he defines his own definition as "c*ontext is the set of environmental states and settings that either determines an application's behavior or in which an application event occurs and is interesting to the user*". Also in 2000, Abowd [17] mentioned that although it is not possible to provide a complete definition of context, it is possible to obtain a good minimal set of necessary context features by responding to "five W's": who, what, where, when, and why. According to Krish [18] "*context is a highly structured amalgam of information, physical and conceptual resources that go beyond the simple facts of who or what is where and when to include the state of digital resources, people concepts and mental state, task state, social relations, and the local work culture, to name a few ingredients*". This definition also applies to the field of mobile computing, moreover, it only focuses on temporal and spatial features and does not consider other dimensions of context. In 2012, Verbert [19] proposes context in recommender systems domain as "an aggregate of various categories that describe the setting in which a recommender is deployed, such as the location, current activity, and available time of the learner". This definition, same as previous, is also described for a specific domain and does not cover other context dimensions. As a result, we claim that there is room for providing a better definition encompassing more aspects and features.

### 2.2 Context Awareness in Big Data Research

In this section, we will investigate context awareness in big data research and compare these studies in different domains as shown in Table 1.

---

[2] https://www.merriam-webster.com/dictionary/context

Table 1: Comparison of context aware big data studies in different domains

| Application Domain | Ref | Goal | Model Proposed | External Dataset | Context Features | Year |
|---|---|---|---|---|---|---|
| The Internet of Things | [20] | Fusion of multidomain context | - | - | Location, Time, User Info, Weather, System Info | 2017 |
| | [21] | Reduction of unnecessary information | - | - | Location, Time, Weather, System Info | 2009 |
| | [22] | representation of Contextual info. in smart city domain | Yes | - | Location, Time, User Info, Weather, System Info, Domain Info | 2018 |
| Mobile based Applications | [23] | Performance maximization | - | - | Location, Time, User Info, System Info | 2018 |
| | [24] | Improved User targeting | - | - | Location, Time | 2017 |
| | [25] | Improved understanding and management of users' behavior | - | - | Location, User Info, Domain Info. | 2014 |
| | [26] | Increased added value of semantic context-aware services | - | - | Location, System Info. | 2013 |
| | [27] | Improved task assignment to the users | - | - | Location, Time, User Info. | 2013 |
| Recommender Systems | [28] | Improved service recommendation to users | - | - | Location, User Info, Domain Info. | 2014 |
| | [29] | Smarter service recommendation to users in smart city | Yes | - | Location, Time, User Info, Weather, System Info, Domain Info. | 2017 |
| | [30] | Improved Geolocation recommendation | - | Weather Underground API | Location, Time, User Info. | 2016 |
| | [31] | Better service Recommendation to users | Yes | - | Location, Time, User Info, Domain Info. | 2016 |
| | [32] | User recommendations based on social interactions | - | Yes * | Location, User Info, Weather, Domain Info. | 2014 |
| | [33] | Presenting a middleware to collect context information | Yes | Yes* | Location, User Info. | 2014 |
| | [34] | facilitation of social networking | - | - | Location, User Info | 2013 |
| | [35] | Providing a reusable location-based group management service | - | - | Location, Time, User Info | 2011 |
| E-health | [36] | Clinical knowledege sharing | - | Yes* | Location, Time, User Info, Domain Info | 2017 |
| | [37] | facilitation of cooperation between several health care partners and patients | Yes | - | User Info, Domain Specific Info, System Info. | 2016 |
| | [38] | Clinical knowledege sharing | Yes | - | Location, Time, User Info, Domain Info. | 2015 |
| Transportation | [39] | Route Optimization | - | - | Time, Domain Specific Info. | 2013 |

* Generally stated and no specific datasets are mentioned

As shown in Table 1, there are five different domains in which context-awareness is considered in big data applications. These include internet of things, mobile computing, recommender systems, e-health, and transportation. Out of these 21 studies, only 6 papers have presented a specific context model for their domain and the rest have only used contextual features in their work. By examining the studies listed in Table 1, we found that only four papers have used external data sources for their purpose some of which have used spatiotemporal context features [30], [33] while [32] has only used social networks information as external data sources.

Some features such as location, time, and user information are used as context features in all domains. On the other hand, some features, such as agent context in e-health and vehicle speed in transportation, are employed only in one domain, which are domain specific information.

In the internet of things and mobile computing domains, in addition to mutual features, system context features such as device type and communication cost are more frequently used. Given the limitations such as limited energy and processing resources in these domain, using information about the system and its context can help the quality management process. On the other hand, in domains of recommender systems and transportations, in addition to common features, climatic variables such as weather, humidity, and pollution are more frequently used to offer higher quality recommendations. Organization context such as agent and group context have also been used in the e-health domain, where this information can improve interpretation of the data value.

Based on the review of the context-aware works presented in Table 1, it can be concluded that not only are the models presented in the big data domain limited to using the information that is appropriate for describing the amount of data (such as environmental and system information), but also the studies which have used context features to enhance their performance. However, we believe that all the information related to the discovery and extraction method of contextual data as well as the data structure and any information that can facilitate big data quality management should be considered in addition to previous features. Therefore, there is a need for a context classification framework that considers all dimensions of context for big data quality.

## 3  PROPOSED FRAMEWORK

As described in Section 2.1, different definitions of context are used by researchers in different domains. It can be seen that the context definitions are either general or provided for a specific domain. None of the definitions outlined can satisfy our needs for getting comprehensive information that need to be considered in the big data quality management process. Therefore, before introducing out proposed context framework, we need a new definition of context that supports the new dimensions of information needed for big data quality. So, we provide the following definition for context:

*Context is any information that can be obtained either directly from the data processing environment such as historical data, or indirectly from other sources (such as ontology, web services, and social network) that provides a proper interpretation of the data value and facilitates the process of big data quality management*.

The idea behind this definition is that to collect contextual information regarding big data quality, four important research questions need to be answered.
RQ 1. What kind of context is available?
RQ2. From what sources is the context taken and what data structure does it have?
RQ 3. How is context data collected and processed?
RQ 4. Which quality factors are affected by the context data?

By answering these four questions, the context needed for the process of the big data quality management is achieved.

During the last decade there have been some efforts to use contextual information for quality assessment of big data. These studies differ from each other in terms of the contextual variables, context data types, and external data sources they have used. In this section, we will present our proposed framework which classifies the state of the art in four main dimensions, including context category, context source, discovery and extraction method, and the quality factors affected by that context data.

In order to develop our classification framework for big data quality, first the big data studies are reviewed and those using context information are selected. These studies are examined more precisely and the context models are extracted along with each of the constituent variables/features to be considered for presenting the final model. Additionally, context-aware studies in the reference paper [9] are also selected. The context features used in these studies have also been extracted to integrate with previous results and to provide the final context model for big data quality.

In the following, we describe the presented context classification framework for big data quality, which is depicted in Figure 1.

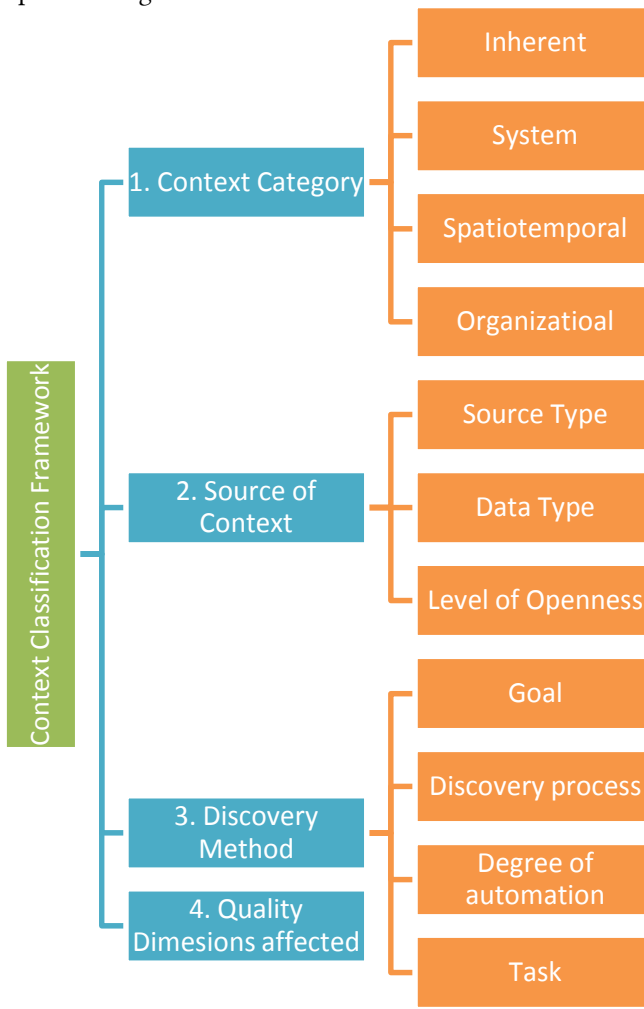

Figure 1: Context classification framework for big data quality

The purpose of the proposed model is to classify the relevant contexts for big data quality management process and as mentioned before, the proposed context classification framework can answer the four questions mentioned.

## 3.1 Context Category

Given the context category, the first question, RQ 1, can be answered. We distinguish four types of context categories: (a) inherent context, (b) system context, (c) spatiotemporal context, and (d) organizational context. These types of context are defined below.

- **Inherent Context** includes any information about data value and the inherent characteristics of the data, such as data schema and constraints. Considering a number of sensors monitoring vital signs of a patient such as blood pressure, to assess the quality of the received data, information including the valid thresholds for minimum and maximum blood pressure can be retrieved from the data schema, or other sources.
- **System Context** describes all information about data source and relevant infrastructure. This includes features and limitations of hardware or software such as power consumption, device type and memory capacity. Knowing these features and limitations of the system can help to assess data quality. For example, checking the value produced by a sensor is much easier with knowing the battery level of sensor, since if we consider the reliability level of a sensor in the assessment process, we can estimate its maximum accuracy to make it easier to evaluate the quality of the data produced.
- **Spatiotemporal Context** addresses any temporal or spatial information that is related to the data value. There are attributes such as time and location that determine when and where data is generated. These can be used to get other useful information, including the weather conditions, nearby resources, temperature, humidity. These information can be extracted from external sources like ontologies, websites, social networks and services. Considering a set of sensors responsible for reading temperature of a forest. It is possible that some of these sensors produce values indicating a high temperature. Having access to additional data source, for instance a web service that provides weather condition in nearby locations, it can be better determined whether the data value is correct or some other factor (i.e. lightning) has caused a disruption to the sensor producing those values.
- **Organizational Context** defines features, limitations, specifications, and any information that are provided directly from relevant organizations, or indirectly by other sources such as websites, ontologies, and social networks, which are responsible for big data quality management. Sometimes some organizational constraints lead to lower quality of data being produced. Suppose that a limited-budget organization plans to process its big data in the cloud environment and evaluate its quality. The low budget has led to the leasing of non-ideal computational platform, and also storage constraints have led to reduction in the data volume, and finally this has led to a decline in the data quality. Moreover, information about employee roles and skills can help to interpret the quality of data, since the low data quality may be due to the inappropriate behavior of a user who lacks the required proficiency with the tasks and has led to the production of poor data or it might be due to another organizational reason.

## 3.2 Source of Context

Among the information that has been overlooked in the related works is the source of context, which includes source type, contextual data types, and the level of openness. These features, which are described next, can be used to answer the second question (RQ 2).

- **Source Type:** The source type can be examined from two perspectives. In the first one, it is determined whether the source type is internal or external. Any information that can be obtained directly from the data source (such as data values and schemas) is considered under the umbrella of internal context, and any information obtained from other sources (such as ontology, websites, or social networks) that provides a proper interpretation of the context category (data value, data source and organization concerned) is called external context. From the second point of view, the source type can be domain specific or general. For example, in [39], variables such as speed and cost of energy are used which are domain specific (i.e. location-based services) but in many studies, context variables such as location or time are used which are general and can be used in different domains.
- **Contextual Data Types:** As mentioned in [40], the data in the big data domain can structured, semi-structured, and unstructured. In structured data types, such as relational tables, data can be stored and retrieved in databases, and processed in fixed predefined formats. Unstructured data, such as text and multi-media, refers to the data that have no specific structure and this makes it much harder to process such data than structured data. Semi-structured data such as streaming data, contains both of the above definitions and although they do not have a specific and fixed schema, they contain vital tags that separate information.
- **Level of Openness:** In order to use external context, one of the important factors to consider is the level of openness. There are several definitions of openness, among all, the most cited is [41], which defined open access as "*access on equal terms for the international research community at the lowest possible cost, preferably at no more than the marginal cost of dissemination. Open access should aim at being easy, timely, user-friendly and preferably Internet-based*". Also in [42], level of openness is classified based on Tim Berners-Lee scheme. Tim Berners-Lee, introduced the concept of a "5-star deployment scheme" for the first time, and proposed a score in the scale of 1 to 5 for each published data[43]. If the data is published on the web, it will get one star. Further, it also receives two stars if the published data is in a machine-readable format (for instance not in a scanned image format). Moreover, if it is available in a non-proprietary format (i.e. csv), it gets three stars. To get the fourth star, in addition to the previous requirements, data needs to be published using the open standard formats provided by w3c, like RDF. Finally, if data meets all the above requirements and it is linked with any other open data, it will receive five stars. By knowing the level of openness, it can be ascertained whether the context was obtained from an external open data source or from a private organization.

### 3.3 Discovery Method

The contextual information needed for the big data quality management is not limited to context category or source of context and it includes information on the discovery and extraction method of contextual data, including goal, discovery process, degree of automation, and task. Using these features, it is possible to provide an answer for the third question (RQ 3). Next, these features are introduced.

- **Goal:** Different methods have been introduced in the big data quality literature for different purposes, including profiling, assessing, monitoring, and improving the quality of data. Some methods [44] identify the data constraints by examining the data and obtaining the necessary information by data profiling, while others [45], seek to assess the quality of data using a set of predefined metrics, such as accuracy and completeness. Additionally, in some works [46], data is monitored, for example using outlier detection methods, to identify and report the faulty data. Furthermore, some researchers [47] make a step forward and try to improve data quality after determining the low quality data.
- **Discovery Process:** Another information that is important about the discovery and extraction method is the type of discovery process that has been employed. It can be online, offline, or hybrid. In an online process [48], data are continuously generated (e.g. data stream) and big data quality management is also conducted continuously without interruption. On the other hand, in offline methods [49], [50], data are stored in a repository and quality management is done statically on the stored data in a batch manner. In addition, in the hybrid [51], [52] methods, both types of online and offline techniques can be used. Actually, while the evaluation process is continuous, it also uses stored data to improve performance.
- **Degree of Automation:** The degree of automation is also one of the information obtained from the discovery method, which can be either automated or semi-automated. Automated (unsupervised) methods [53] usually use existing or extended tools to perform big data quality analysis without any intermediary human intervention. In semi-automated (supervised) methods [10], part of the work is performed or decided by the user or an expert.
- **Technique:** There are several techniques that can be helpful for improving big data quality management, each of which has its own characteristics and limitations. Learning based [47], model based [54], rule based [55] and distance based [56] methods are among the techniques employed in the literature for the purpose of executing different tasks in big data quality management. For example, an important characteristic of the distance-based methods is that they entail high computational cost, which makes them inappropriate for processing high volumes of data and hence, can lead to poor management of data quality.

## 3.4 Quality Factors Affected

The last question that needs to be answered is what quality factors are affected by a specific context feature (RQ 4). ISO [57] has pointed out fifteen data quality dimensions, however, some of these have not been applied in the big data studies. In [58], it is stated that some quality dimensions are widely used in the big data quality studies, including availability, usability, reliability, relevance, and presentation quality. All these quality dimensions and their elements are defined in [58] which are explained below:
- **Availability:** Means the level of convenience of users to get relevant contextual information about data. Accessibility, authorization, and timeliness are three elements of availability factor. Accessibility is defined as the degree in which the users are able to obtain the required information easily. This element has a direct relationship with the level of openness. The more data available, the greater is the level of openness of data. Authorization means providing the users with the required privileges to use data and information, and timeliness is the time delay from data collection to operation. Timeliness is important for the processing of big data, since generally, time is very critical and data values can be changed quickly at any moment.
- **Usability:** If data and related information meet the users' needs, it is usable. User needs may be obtained from documentations or metadata of a system under study. Credibility, documentation, and metadata are elements of the usability dimension. Credibility is defined as the degree that the user believes the data is correct and in accordance with reality in a particular context of use. Documentations contain information such as definitions, limitations and rules that can be used in the big data quality management process. The meaning of data varies with the variation of data sources, and there might be different interpretations of data with the same name, in which case there is no other way than to use metadata to understand the obtained information.

- **Reliability:** Reliability deals with the question of whether data can be trusted. Accuracy, completeness, and consistency are some quality factors that determine reliability. The closeness of the given data value to the actual data is called accuracy. In some cases, it is easy to determine the accuracy of data, but in other cases, additional information is needed to describe the domain context. Completeness indicates whether there is a gap between what is expected to be collected and what is actually collected. For example, if a data repository does not include definition of a color attribute, the color of an object of interest cannot be stored in the repository and hence, any quality management task that requires analysis of this attribute fails. Consistency refers to the logical relationship of the data values associated with each other. For example, the value of a person's birth date attribute should not conflict with the value of his/her age attribute.
- **Relevance:** Refers to the degree of relation between obtained information and users' demands in a specific quality analysis task. Fitness is an element of relevance factor, which is defined as the degree of data generated that matches the users' needs.
- **Presentation Quality:** This determines how the data is described and how easily the users understand the information. Readability is an element of presentation quality that is described as the ability of data information to be correctly understood and explained by the users.

Each of the above data quality factors are among the dimensions that can be affected by the incoming data information, which are necessary for big data quality management. By evaluating these quality factors in big data, the quality of the given data will be determined.

## 4 COMPARING STATE OF THE ART IN THE PROPOSED FRAMEWORK

In Section 3, we introduced a framework for classifying context for quality assessment of big data and described each dimension along with the sub-dimensions and potential values. Here, we are going to compare the state of the art using our proposed model as shown in Table 2. The rows of this table show the dimensions and sub-dimensions of our framework. As the table shows, some major rows indicating dimensions of the framework are divided into minor rows indicating sub-dimensions (such as inherent, system, source type and goal).

The columns in Table 2 indicate the works conducted in the area of big data quality which have used context features. These works have been carefully studied and inspected in our systematic review [9] to achieve a comprehensive context model.

As indicated in Table 2, none of the studies has covered the inherent and system context features. On the other hand, only one paper used organizational features. Although spatiotemporal features have been used in all papers, there is no diversity of context feature selections and most studies are limited to using time and location for their purpose, while only two studies have used additional features like weather conditions to improve their performance.

Furthermore, six studies have used the internal data source, while only in [54], weather context is taken into consideration by retrieving weather information from the meteorological public website. Another interesting observation from the analysis of the studies is that most studies (57%) use inaccessible datasets (got zero star) in order to evaluate their work, which reduces the level of openness of data and makes it difficult to reproduce their experiments and results. With the expansion of data sources and the development of some domains such as IoT, most of the analysis is done on the streaming data type. This is evidenced from Table 2 that the data type of all papers is data stream, which emphasizes the importance of this type of data in the big data era.

Most studies (57%) processed data for the purpose of improving data quality, while one paper focuses on quality evaluation and two other papers address monitoring quality of data. Out of these methods, in four papers the processing method is performed online and in a continuous mode. In addition, discovery process in two papers is hybrid and in one paper this is performed through offline process on stored data. Also, four papers presented automatic methods which do not require any user intervention, while in three studies, the user has a specific role in the big data quality management process. By inspecting the studies in Table 2, it can be observed that a variety of techniques have been used to increase the performance of the quality management method, including sampling, learning based, model based and distance based.

By examining the quality dimensions used in the studies, it is shown that only in one paper, which aims at evaluating quality of data, data quality dimensions have been examined and in other studies this has been neglected.

Table 2: Comparison of studies in the proposed framework

| | Dimensions | [10] | [54] | [59] | [60] | [61] | [62] | [63] |
|---|---|---|---|---|---|---|---|---|
| Context Category | Inherent | - | - | - | - | - | - | - |
| | System | - | - | - | - | - | - | - |
| | Spatiotemporal | Time | Location | Location, Time, Weather | Location, Time, Weather | Location, Neighbors | Location, Time | Location, Coverage area |
| | Organizational | Budget | - | - | - | - | - | - |
| Source of Context | Source Type | Internal | Int & Ext (Wind speed) | Internal | Internal | Internal | Internal | Internal |
| | Data Type | Stream | Stream | Stream | Stream | Stream | Stream | Stream |
| | Level of Openness | 0 | 3 stars | 0 | 0 | 2 stars | 3 stars | 0 |
| Discovery Method | Goal | Assessment | Improvement | Monitoring | Monitoring | Improvement | Improvement | Improvement |
| | Discovery Process | Offline | Online | Hybrid | Hybrid | Online | Online | Online |
| | Degree of Automation | Semi-Automate | Semi-Automate | Automated | Automated | Automated | Automated | Semi-Automate |
| | Technique | Sampling | Model based | Learning based | Learning based | Distance based | Learning based | Distance based |
| Quality Dimensions | | Availability, Reliability | N/A | N/A | N/A | N/A | N/A | N/A |

## 5 EVALUATION

In this section, the evaluation of the proposed context model is explained. There exist different strategies in the literature for the purpose of evaluating a conceptual model. In section 5.1, we first describe these evaluation strategies, and then in section 5.2, the evaluation of the proposed model is explained.

### 5.1 Evaluation Strategies

Conceptual models have been evaluated in different studies in various ways. In [64] evaluating strategies of conceptual models are reviewed, which broadly speaking, can be divided into two categories: empirical and non-empirical.

In empirical techniques, the evaluation is performed through conducting a survey, laboratory experiment, or case study. Surveys are a good way to evaluate information modeling methods, which gather all information of methods, including opinions, beliefs, impressions on models, advantages, disadvantages, and any related context that can be used to compare methods. In laboratory experiment method, the models are evaluated based on determining the independent variables such as accuracy and cost (in terms of modeling time). Paper [65] has used laboratory experiment to study the effect of modeling construct. Moreover in case study method, which is commonly used in data sciences [29], [36], researchers investigate an individual or group with its real-life context. For this reason, the richness of this method is higher than the previous two methods.

On the other hand, in non-empirical techniques, feature comparison, meta modeling, metrics-based, paradigmatic analysis, contingency identification, ontological evaluation, and approaches based on cognitive psychology have been employed. The main idea behind the feature comparison technique is to use different methods to model a same domain, and see how the different models tackle the same problem.

The researchers of [66] have evaluated different programming methods with this technique. Using meta models is another way of comparing methods, and the idea of this technique is to identify similar parts in different models and try to evaluate models by structurally investigating analogies between them. For instance, [67] compare various software design methods with this technique.

Some researchers have analyzed the assumptions behind system development and introduced another method of evaluation, called paradigmatic analysis. Here, the researchers express their views on a problem and compare it with those of the other approaches. This method has been used in [68] for analyzing information systems techniques.

In contingency identification method, researchers [69] that compare different information system techniques, evaluate the methods based on the degree of risk they have. In these methods, after identifying the problems of the methods, a less risky model is selected.

Also in ontological evaluation technique, different model constructs are mapped to ontological constructs in order to evaluate the quality of the models by identifying construct overload, construct redundancy, construct excess, and construct deficit. Ontological evaluation method is used in [70] for systems analysis and design methods, and in [71] for information systems.

Approaches based on cognitive psychology are another evaluation technique. Since modeling methods collect knowledge of a domain, it is necessary to examine the cognitive aspect of modeling. So, in some studies [72], [73], which are in information modeling era, the impact of cognitive psychology on statements that may be derived from the use of modeling methods, is examined.

Finally in this technique, the models are compared based on a set of predefined metrics (e.g. [74] in object oriented methods and [75] in systems development methods). The metrics can provide a valuable aid in evaluating the model, as well as in different domains and applications, different metrics are used, and the purpose of selecting these metrics is to evaluate the complexity and appropriateness of the model.

## 5.2 Evaluation of Proposed Model

Among the methods mentioned in the previous section, the metrics approach and feature comparison are chosen to evaluate the proposed context classification framework and compare it with other models.

*5.2.1 Evaluation based on Feature Comparison.* As mentioned in Section 5.1, one of the evaluation strategies is feature comparison, which is used to evaluate proposed model with other context aware models. Table 3, compares all presented context models in the big data domain, with the proposed context framework.

Table 3: Comparative Evaluation

| Ref | Year | Type of the Model | Evaluation Method | Context Dimensions | | | | Domain |
| --- | --- | --- | --- | --- | --- | --- | --- | --- |
| | | | | Context Category | Source of Context | Discovery Method | Quality Dimensions Affected | |
| [33] | 2014 | Multi-dimensional | Case Study | ✓ | - | - | - | Recommender Systems |
| [38] | 2015 | Hierarchical | Discussion | ✓ | - | - | - | E-health |
| [31] | 2016 | Multi-dimensional | Discussion | ✓ | - | - | - | Recommender Systems |
| [37] | 2016 | Layered | - | ✓ | - | - | - | E-health |
| [29] | 2017 | Multi-dimensional | Case Study | ✓ | - | - | - | Recommender Systems |
| [22] | 2018 | Tabular | - | ✓ | - | - | - | The Internet of Things |
| Proposed Framework | 2019 | Hierarchical | Theoretically | ✓ | ✓ | ✓ | ✓ | Big Data Quality |

Among the papers that have presented a model, studies [29], [31], [33] proposed a multidimensional model, [37] uses a layered model, [38] uses a hierarchical, and [22] uses a tabular model. Also in the proposed framework, due to the readability of hierarchical models, this type of model is used. On the other hand, out of these six studies, only four papers have evaluated their proposed model. In [29], [33], [36] case study method have been used and in [31], [38] authors evaluate their model based on discussion of its characteristics, without any experimental evaluation. Of all the evaluation methods outlined in Section 5.1, the proposed context framework is evaluated by metrics approach and feature comparison methods, which are theoretical approaches. In terms of the number of context dimensions covered, all models cover only the context category dimension, while the proposed method considers all contextual information, including context category, source of context, discovery and extraction method, and quality dimensions affected by. As can be seen from Table 3, only the domain of the proposed framework is big data quality and other models are presented in other big data domains, such as the Internet of things, e-health, and recommender systems.

Therefore, by comparing the proposed framework with other models presented in the big data domain, it can be concluded that the proposed method, although more readable because of the hierarchical model type, it covers more contextual dimensions than the other models.

*5.2.2 Evaluation based on Metrics Approach.* In this part, our context classification framework is evaluated based on metrics approach. Therefore, understandability, comprehensiveness, usefulness, and richness are used as selected metrics to compare the proposed model with other methods.

❖ **Understandability**

To evaluate the model, one of the metrics is understandability. It refers to the degree in which the user understands the content of the model correctly and simply. Different studies have examined the impact of understandability on different classification models [76], [77]. As shown in Table 3, only one model [38], has a hierarchical structure while the others are either layered or multidimensional. To increase understandability, our proposed model has been presented in hierarchical structure.

❖ **Comprehensiveness**

Another metric for evaluating the proposed context model is comprehensiveness, in which the models are compared based on their content to determine which model has a higher coverage over the set of concepts in its domain. Thus, different models are compared based on their structural similarities. By studying the previous models, we tried to cover all aspects of contextual information. As presented in Table 3, all models have only considered context category, and other aspects of context data such as data source, discovery method, and quality dimensions have not been taken into account. Therefore, from the content point of view, the proposed model is more comprehensive than the other models.

❖ **Usefulness**

Another criterion that can be used to evaluate the proposed model is usefulness. According to Abowd's definition, a good minimal set of necessary context information is achieved by responding to "five W's": what, who, where, when, and why [17]. Here, we define each of these five W's and explain how our proposed model can address all.

- **What** is defined as "*the interaction in current systems either assumes what the user is doing or leaves the question open*" for his domain (i.e. ubiquitous computing) [17]. In the domain of big data quality, "what" refers to the information about the generated data value. So the questions that arise from "what" are: 1) What data value is generated? and 2) What data/source types does it have? The proposed model is able to answer to these question by considering the **inherent** dimension of **context categorization** and **data type** of **source of context**. Many of the related models have answered the first question with regard to the data value, but for the second question they have not considered any context dimension.
- **Who** is defined as "current systems focus their interaction on the identity of one particular user, rarely incorporating identity information about other people in the environment." [17]. In the context of big data quality, "who" defines as the data source itself as well as any information about the data source. Questions such as 1) What are the features and limitations of the data source? 2) Is the context coming from an open/external source? 3) What is the level of openness? Previous models can only answer the first question by obtaining system context, while the proposed model is able to answer all three questions with considering both **system** dimension of **context categorization** and **source type** of **source of context**.
- **Where and When:** The meaning of "where" and "when" suggests that it is needed to consider the time and location associated with the data, which has a similar definition to [17]. All previous models have considered time and location features as context, For example, weather conditions are one of the essential contexts which is considered in models [22], [29]. On the other hand, in these two papers weather variables are available with internal (sensor-based) sources, while in the proposed model, it is possible to obtain information from external sources such as ontology, websites and social networks with considering **spatiotemporal** dimension in context category and **internal/external source type** in **source of context**.
- **Why:** "*Why the person is doing something?*" in the big data quality domain, the last question that needs to be answered is why should we manage big data quality? According to Abowd, "why" is the most challenging question that needs to be answered. Previous models have not answered this question and the proposed model is able to answer the question by considering the **organizational** dimension. Based on [9], organizations have concluded that big data processing has a significant impact on their business. On the other hand, poor data quality has led them to make the wrong decisions. As a result, organizations have sought to assess the quality of data to improve their business. So, this question can be answered based on organizational needs, and proposed model can answer the question "why" by considering the **organizational** information in **context category**.

- ❖ **Richness**

The last metric for evaluating the proposed context model is its richness, which is determined by comparing the context features used in this model and other existing models. The more context features used in the model, the richer the model is expected to be. As already discussed, different model features are shown in Tables 1 and 2. Compared to other models, the proposed model has unique contextual features including organizations' context (e.g. budget), discovery and extraction method context (e.g. goal and discovery process) and any information which can be obtained from external resources such as ontologies, websites, and social networks that have not been used in previous models. As noted earlier, organizational constraints, one of which is the role of individuals in the organization, may be the reason for the poor quality. Suppose in an organization, a person is processing and analyzing data and, for a variety of reasons, such as fatigue, inexperience, or unfamiliarity with his or her task, can produce low quality data. Knowing the various factors that lead to the production of poor data can be exploited, and this type of context is not considered in other models. So, by comparing the features of the proposed model with other models, it can be concluded that the proposed model is richer and is able to support more analysis tasks that require different types of contextual data. Based on the metrics approach, it can be claimed that the proposed context model is more understandable, more comprehensive, more useful and richer than other related models.

## 6   CONCLUSION

Several context-aware studies have been presented in the field of big data quality, most of which use spatiotemporal information to enhance the performance of their work, while more information can be used for this purpose in big data applications. This paper presents a context classification framework for big data quality that classify the context features into four primary dimensions, including context category, source of context, discovery and extraction methods, and quality factors affected by. Through this classification, it is possible to identify many context features that influence the big data quality management process. The proposed model is evaluated using the metrics-based approach and also through feature comparison, which are among the conceptual model evaluation methods. The results demonstrate that the proposed model is more understandable, richer, more comprehensive and more useful than other proposed models in the big data domain.

## REFERENCES

[1]   H. Liu, F. Huang, H. Li, W. Liu, and T. Wang, "A Big Data Framework for Electric Power Data Quality Assessment," in *2017 14th Web Information Systems and Applications Conference (WISA)*, 2017, pp. 289–292.
[2]   S. E. Bibri, "Big Data and Context – Aware Computing Applications for Smart Sustainable Cities," no. November, 2016.
[3]   P. Malik, "Governing Big Data: Principles and practices," *IBM J. Res. Dev.*, vol. 57, no. 3/4, p. 1:1-1:13, May 2013.
[4]   S. Xie and Z. Chen, "Anomaly Detection and Redundancy Elimination of Big Sensor Data in Internet of Things," Mar. 2017.
[5]   X. Dong, H. He, C. Li, Y. Liu, and H. Xiong, "Scene-Based Big Data Quality Management Framework," Springer, Singapore, 2018, pp. 122–139.
[6]   N. Abdullah, S. A. Ismail, S. Sophiayati, and S. M. Sam, "Data quality in big data: A review," *Int. J. Adv. Soft Comput. its Appl.*, vol. 7, no. Specialissue3, pp. 16–27, 2015.
[7]   J. Taylor, "IBM big data and information management," 2011.

[8] Thomas.C.Redman, *Getting in front of Data : The who does what.* Technics Publication, 2016.
[9] M. Mirzaie, B. Behkamal, and S. Paydar, "Big Data Quality: A systematic literature review and future research directions," Apr. 2019.
[10] D. Ardagna, C. Cappiello, W. Samá, and M. Vitali, "Context-aware data quality assessment for big data," *Futur. Gener. Comput. Syst.*, vol. 89, pp. 548–562, Dec. 2018.
[11] M. A. Langouri, Z. Zheng, F. Chiang, L. Golab, and J. Szlichta, "Contextual data cleaning," *Proc. - IEEE 34th Int. Conf. Data Eng. Work. ICDEW 2018*, pp. 21–24, 2018.
[12] B. N. Schilit and M. M. Theimer, "Disseminating active map information to mobile hosts," *IEEE Netw.*, vol. 8, no. 5, pp. 22–32, Sep. 1994.
[13] B. Schilit, N. Adams, and R. Want, "Context-Aware Computing Applications," *1994 First Work. Mob. Comput. Syst. Appl.*, pp. 85–90, 1994.
[14] A. Schmidt, K. A. Aidoo, A. Takaluoma, U. Tuomela, K. Van Laerhoven, and W. Van de Velde, "Advanced Interaction in Context," Springer, Berlin, Heidelberg, 1999, pp. 89–101.
[15] G. D. Abowd, A. K. Dey, P. J. Brown, N. Davies, M. Smith, and P. Steggles, "Towards a Better Understanding of Context and Context-Awareness BT - Handheld and Ubiquitous Computing: First International Symposium, HUC'99 Karlsruhe, Germany, September 27–29, 1999 Proceedings," pp. 304–307, 1999.
[16] G. Chen and D. Kotz, "A Survey of Context-Aware Mobile Computing Research," *Dartmouth Comput. Sci. Tech. Rep. TR2000-381*, Nov. 2000.
[17] G. D. Abowd and E. D. Mynatt, "Charting Past , Present , and Future Research in Ubiquitous Computing," vol. 7, no. 1, pp. 29–58, 2000.
[18] D. Kirsh, "The Context of Work," *Human–Computer Interact.*, vol. 16, no. 2–4, pp. 305–322, Dec. 2001.
[19] K. Verbert *et al.*, "Context-aware recommender systems for learning: A survey and future challenges," *IEEE Trans. Learn. Technol.*, vol. 5, no. 4, pp. 318–335, 2012.
[20] B. Jia, S. Liu, Y. Guan, W. Li, and W. Ren, "The fusion model of multidomain context information for the internet of things," *Wirel. Commun. Mob. Comput.*, vol. 2017, 2017.
[21] V. Q. Son, B. L. Wenning, A. Timm-Giel, and C. Görg, "A model of wireless sensor networks using context-awareness in logistic applications," *2009 9th Int. Conf. Intell. Transp. Syst. Telecommun. ITST 2009*, pp. 2–7, 2009.
[22] B. Boudaa, S. Hammoudi, and S. M. Benslimane, "Towards an Extensible Context Model for Mobile User in Smart Cities," Springer, Cham, 2018, pp. 498–508.
[23] S. Klos Nee Muller, C. Tekin, M. Van Der Schaar, and A. Klein, "Context-Aware Hierarchical Online Learning for Performance Maximization in Mobile Crowdsourcing," *IEEE/ACM Trans. Netw.*, vol. 26, no. 3, pp. 1334–1347, 2018.
[24] L. B. Sko, "Curious Cat – Mobile , Context-Aware Conversational Crowdsourcing r r r," vol. 35, no. 4, 2017.
[25] A. Ahmad *et al.*, "A framework for crowd-sourced data collection and context-aware services in Hajj and Umrah," *Proc. IEEE/ACS Int. Conf. Comput. Syst. Appl. AICCSA*, vol. 2014, pp. 405–412, 2014.
[26] A. Uzun, L. Lehmann, T. Geismar, and A. Küpper, "Turning the OpenMobileNetwork into a live crowdsourcing platform for semantic context-aware services," p. 89, 2013.
[27] I. Carreras, D. Miorandi, A. Tamilin, E. R. Ssebaggala, and N. Conci, "Matador: Mobile task detector for context-aware crowd-sensing campaigns," *2013 IEEE Int. Conf. Pervasive Comput. Commun. Work. PerCom Work. 2013*, no. March, pp. 212–217, 2013.
[28] D. Dqg *et al.*, "Towards a Mobile and Context-Aware Framework from Crowdsourced Data," in *The 5th International Conference on Information and Communication Technology for The Muslim World (ICT4M)*, 2014.
[29] N. Gutowski, T. Amghar, O. Camp, and S. Hammoudi, "A framework for context-aware service recommendation for mobile users: A focus on mobility in smart cities," *From Data To Decis.*, no. August, 2017.
[30] S. A.El-Moemen, T. Hassan, and A. A.Sewisy, "A Context-Aware Recommender System for Personalized Places in Mobile Applications," *Int. J. Adv. Comput. Sci. Appl.*, vol. 7, no. 3, pp. 442–448, 2016.
[31] N. D. A. B. Navarro, C. A. Da Costa, J. L. V. Barbosa, and R. D. R. Righi, "A context-aware spontaneous mobile social network," *Proc. - 2015 IEEE 12th Int. Conf. Ubiquitous Intell. Comput. 2015 IEEE 12th Int. Conf. Adv. Trust. Comput. 2015 IEEE 15th Int. Conf. Scalable Comput. Commun. 20*, pp. 85–92, 2016.
[32] X. Hu, X. Li, E. C. H. Ngai, V. C. M. Leung, and P. Kruchten, "Multidimensional context-aware social network architecture for mobile crowdsensing," *IEEE Commun. Mag.*, vol. 52, no. 6, pp. 78–87, 2014.
[33] V. Arnaboldi, M. Conti, and F. Delmastro, "CAMEO: A novel context-aware middleware for opportunistic mobile social networks," *Pervasive Mob. Comput.*, vol. 11, pp. 148–167, 2014.
[34] A. Chin, B. Xu, H. Wang, L. Chang, H. Wang, and L. Zhu, "Connecting people through physical proximity and physical resources at a conference," *ACM Trans. Intell. Syst. Technol.*, vol. 4, no. 3, p. 1, 2013.
[35] R. Lubke, D. Schuster, and A. Schill, "MobilisGroups: Location-based group formation in Mobile Social Networks," *2011 IEEE Int. Conf. Pervasive Comput. Commun. Work. PERCOM Work. 2011*, pp. 502–507, 2011.
[36] O. Anya and H. Tawfik, "Designing for practice-based context-awareness in ubiquitous e-health environments," *Comput. Electr. Eng.*, vol. 61, pp. 312–326, 2017.
[37] K. Wan and V. Alagar, "Context-aware, knowledge-intensive, and patient-centric Mobile Health Care Model," *2015 12th Int. Conf. Fuzzy Syst. Knowl. Discov. FSKD 2015*, pp. 2253–2260, 2016.
[38] O. Anya, H. Tawfik, and D. Al-Jumeily, "Context-aware clinical knowledge Sharing in crossboundary e-health: A conceptual model," *Proc. - 15th IEEE Int. Conf. Comput. Inf. Technol. CIT 2015, 14th IEEE Int. Conf. Ubiquitous Comput. Commun. IUCC 2015, 13th IEEE Int. Conf. Dependable, Auton. Se*, pp. 589–595, 2015.
[39] Y. Wang, J. Jiang, and T. Mu, "Context-Aware and Energy-Driven Route Optimization for Fully Electric

[40] A. Rai, "What is Big Data - Characteristics, Types, Benefits and Examples," 2019. [Online]. Available: https://www.upgrad.com/blog/what-is-big-data-types-characteristics-benefits-and-examples/#Structured. [Accessed: 02-Sep-2019].
[41] D. Pilat and Y. Fukasaku, "OECD Principles and Guidelines for Access to Research Data from Public Funding," *Data Sci. J.*, vol. 6, pp. OD4-OD11, 2007.
[42] S. Martin, M. Foulonneau, and S. Turki, "1-5 Stars: Metadata on the Openness Level of Open Data Sets in Europe," *Commun. Comput. Inf. Sci.*, vol. 390 CCIS, no. 2013, pp. 234–245, 2013.
[43] T. Berners-Lee, "5-star Open Data." [Online]. Available: https://5stardata.info/en/. [Accessed: 02-Sep-2019].
[44] C. Cappiello, W. Samá, and M. Vitali, "Quality awareness for a Successful Big Data Exploitation," in *Proceedings of the 22nd International Database Engineering & Applications Symposium on - IDEAS 2018*, 2018, pp. 37–44.
[45] M. Klas, W. Putz, and T. Lutz, "Quality Evaluation for Big Data: A Scalable Assessment Approach and First Evaluation Results," in *2016 Joint Conference of the International Workshop on Software Measurement and the International Conference on Software Process and Product Measurement (IWSM-MENSURA)*, 2016, pp. 115–124.
[46] M. A. Rassam, M. A. Maarof, and A. Zainal, "A distributed anomaly detection model for wireless sensor networks based on the one-class principal component classifier," *Int. J. Sens. Networks*, vol. 27, no. 3, p. 200, 2018.
[47] H. Liu, J. Chen, F. Huang, and H. Li, "An Electric Power Sensor Data Oriented Data Cleaning Solution," in *2017 14th International Symposium on Pervasive Systems, Algorithms and Networks & 2017 11th International Conference on Frontier of Computer Science and Technology & 2017 Third International Symposium of Creative Computing (ISPAN-FCST-ISCC)*, 2017, pp. 430–435.
[48] S. Sadik, L. Gruenwald, and E. Leal, "In pursuit of outliers in multi-dimensional data streams," in *2016 IEEE International Conference on Big Data (Big Data)*, 2016, pp. 512–521.
[49] M. G. Rahman and M. Z. Islam, "Missing value imputation using a fuzzy clustering-based EM approach," *Knowl. Inf. Syst.*, vol. 46, no. 2, pp. 389–422, Feb. 2016.
[50] Z.-Y. Qu, Y.-W. Wang, C. Wang, N. Qu, and J. Yan, "A Data Cleaning Model for Electric Power Big Data Based on Spark Framework," *Int. J. Database Theory Appl.*, vol. 9, no. 3, pp. 137–150, 2016.
[51] V. Pullabhotla and K. P. Supreethi, "Adaptive Pre-processing and Regression of Weather Data," Springer, Singapore, 2017, pp. 9–13.
[52] Y. Tian, P. Michiardi, and M. Vukolic, "Bleach: A Distributed Stream Data Cleaning System," in *2017 IEEE International Congress on Big Data (BigData Congress)*, 2017, pp. 113–120.
[53] T. Zaarour, N. Pavlopoulou, S. Hasan, U. ul Hassan, and E. Curry, "Automatic Anomaly Detection over Sliding Windows," in *Proceedings of the 11th ACM International Conference on Distributed and Event-based Systems - DEBS '17*, 2017, pp. 310–314.
[54] S. Gill, B. Lee, and E. Neto, "Context aware model-based cleaning of data streams," in *2015 26th Irish Signals and Systems Conference (ISSC)*, 2015, pp. 1–6.
[55] L. Lu, H. Cheng, S. Xiong, P. Duan, and Y. Xiao, "Distributed Anomaly Detection Algorithm for Spatio-Temporal Trajectories of Vehicles," in *2017 IEEE International Symposium on Parallel and Distributed Processing with Applications and 2017 IEEE International Conference on Ubiquitous Computing and Communications (ISPA/IUCC)*, 2017, pp. 590–598.
[56] Y. Yan, L. Cao, C. Kulhman, and E. Rundensteiner, "Distributed Local Outlier Detection in Big Data," in *Proceedings of the 23rd ACM SIGKDD International Conference on Knowledge Discovery and Data Mining - KDD '17*, 2017, pp. 1225–1234.
[57] ISO, "ISO 25012." [Online]. Available: https://iso25000.com/index.php/en/iso-25000-standards/iso-25012. [Accessed: 02-Sep-2019].
[58] L. Cai and Y. Zhu, "The Challenges of Data Quality and Data Quality Assessment in the Big Data Era," *Data Sci. J.*, vol. 14, no. 0, p. 2, May 2015.
[59] M. A. Hayes and M. A. Capretz, "Contextual anomaly detection framework for big sensor data," *J. Big Data*, vol. 2, no. 1, p. 2, Dec. 2015.
[60] M. A. Hayes and M. A. M. Capretz, "Contextual Anomaly Detection in Big Sensor Data," in *2014 IEEE International Congress on Big Data*, 2014, pp. 64–71.
[61] Y. Zhang, C. Szabo, and Q. Z. Sheng, "Cleaning Environmental Sensing Data Streams Based on Individual Sensor Reliability," Springer, Cham, 2014, pp. 405–414.
[62] V. Iyer, "Ensemble Stream Model for Data-Cleaning in Sensor Networks," 2013.
[63] S. Pumpichet, "NOVEL ONLINE DATA CLEANING PROTOCOLS FOR DATA STREAMS IN TRAJECTORY , WIRELESS SENSOR NETWORKS A dissertation submitted in partial fulfillment of the requirements for the degree of DOCTOR OF PHILOSOPHY in ELECTRICAL ENGINEERING by Sitthapo," 2013.
[64] K. Siau and M. Rossi, "Evaluation of Information Modeling Methods -- A Review," pp. 1–9, 2004.
[65] D. Batra and J. G. Davis, "Conceptual data modelling in database design: similarities and differences between expert and novice designers," *Int. J. Man. Mach. Stud.*, vol. 37, no. 1, pp. 83–101, Jul. 1992.
[66] S. B. Yadav, R. R. Bravoco, A. T. Chatfield, and T. M. Rajkumar, "Comparison of analysis techniques for information requirement determination," *Commun. ACM*, vol. 31, no. 9, pp. 1090–1097, 1988.
[67] Xiping Song and L. J. Osterweil, "Experience with an approach to comparing software design methodologies," *IEEE Trans. Softw. Eng.*, vol. 20, no. 5, pp. 364–384, May 1994.
[68] J. Iivari and R. Hirschheim, "Analyzing information systems development: A comparison and analysis of eight is development approaches," *Inf. Syst.*, vol. 21, no. 7, pp. 551–575, Nov. 1996.
[69] M. Schipper and S. M. M. Joosten, "A validation procedure for information systems modeling techniques." 25-Feb-1996.

[70] Y. Wand and R. Weber, "An ontological evaluation of systems analysis and design methods - UQ eSpace," in *Proceedings of the IFIP WG 8.1 Working Conference on Information Systems Concepts*, 1989.
[71] Y. Wand and R. Weber, "An ontological analysis of some fundamental information systems concepts," *Proc. Ninth Int. Conf. Inf. Syst.*, vol. 1988, pp. 213–226, 1988.
[72] K. Siau, "Using GOMS for Evaluating Information Modeling Methods," in *Second CAiSWIFIP8. I International Workshop on Evaluation of Modeling Methods in Systems Analysis and Design (EMMSAD'97)*, 1997.
[73] K. Siau, Y. Wand, and I. Benbasat, "Evaluating Information Modeling Methods -- A Cognitive Perspective," in *Workshop on Evaluation of Modeling Methods in Systems Analysis and Design (EMMSAD'96)*, 1996.
[74] M. Rossi, "Evolution of OO Methods: the unified case," *Australas. J. Inf. Syst.*, vol. 4, no. 2, May 1997.
[75] M. Rossi and S. Brinkkemper, "Complexity metrics for systems development methods and techniques," *Inf. Syst.*, vol. 21, no. 2, pp. 209–227, Apr. 1996.
[76] H. Allahyari and N. Lavesson, "User-oriented assessment of classification model understandability," *Front. Artif. Intell. Appl.*, vol. 227, pp. 11–19, 2011.
[77] A. A. Freitas, "Comprehensible classification models," *ACM SIGKDD Explor. Newsl.*, vol. 15, no. 1, pp. 1–10, 2014.